# Control of surface induced phase separation in immiscible semiconductor alloy core-shell nanowires

M. Arjmand[1], J.H. Ke[2], and I. Szlufarska[1, 2]

[1]*Department of Engineering Physics, University of Wisconsin – Madison, Madison, 53706, USA*
[2]*Department of Materials Science and Engineering, University of Wisconsin – Madison, Madison, 53706, USA*

**Abstract**
Semiconductor nanowires have been shown to exhibit novel optoelectronic properties with respect to bulk specimens made of the same material. However, if a semiconductor alloy has a miscibility gap in its phase diagram, at equilibrium it will phase separate, leading to deterioration of the aforementioned properties. One way to prevent this separation is to grow the material at low temperatures and therefore to suppress kinetics. Such growth often needs to be followed by high-temperature annealing in order to rid the system of undesirable growth-induced defects. In this study, we propose a method to control phase separation in core-shell nanowires during high temperature annealing by tailoring geometry and strain. Using a phase field model we determined that phase separation in nanowires begins at the free surface and propagates into the bulk. We discovered that including a thin shell around the core delays the phase separation whereas a thick shell suppresses the separation almost entirely.

## 1. Introduction

High surface to volume ratio and the possibility of a lateral strain relaxation make nanowires promising candidates for growth of semiconductor structures. Semiconductor nanowires have shown outstanding electronic and optical properties, and therefore are being considered for use as lasers [1], light emitting diodes [1,2], transistors [3] and sensors [4]. In particular, many studies have focused on nanowires made of III-V semiconductors because the band gap energy in these materials can be controlled by alloy composition [5,6]. One challenge in growing multi-component heterostructures, such as InGaAs, AlGaAs, and GaAsSb, is that these alloys have a miscibility gap in their phase diagrams and therefore at equilibrium these alloys phase separate [7]. During growth process, phase separation can be kinetically inhibited for most III-V semiconductors because of the relatively low growth temperatures. However, these materials often need to be subsequently annealed in order to remove defects introduced during growth and the high temperature annealing can lead to undesirable phase separation. For instance, Luna *et al.* [8] found spontaneous formation of a lateral composition modulation (LCM) in GaAsBi epilayers grown by Molecular Beam Epitaxy (MBE). Hsieh *et al.* [9] observed LCM in AlGaAs film upon annealing and found that the phase separation was more pronounced near the free surface. The authors proposed a stress-driven vacancy-assisted mechanism to be responsible for this phenomenon. Tang *et al* [10] used linear stability theory to study the role of free surfaces in spontaneous phase separation of alloys in thin films and found that stress relaxation begins at the surface. However, Tang *et al.* did not study the effects of geometry and strain on formation of surface induced compositional modulations. Also they did not investigate potential pathways for controlling this phase separation.

As shown experimentally, the existence of a miscibility gap in phase diagrams of III-V semiconductors leads to compositional modulation during growth and annealing [8,9]. Here, we propose a method to control such phase separation under conditions where phase separation is thermodynamically favorable and kinetically allowed. This method takes advantage of the strain induced by a core-shell geometry of a nanowire. The effects of different factors, such as the miscibility gap, surface and bulk diffusion, and elastic strains, on the compositional modulation in the nanowire structure are investigated using the phase field model.

## 2. Model

We model the annealing process of the nanowires made of a generic immiscible alloy using phase field method combined with elasticity governing equations. We developed this continuum-based model earlier to study the growth of thin films on patterned substrates [11] and continuum approaches have been found to be applicable for strain and stress field calculations in core-shell nanowires [12]. In our model we do not account for possible faceting of the nanowires because our goal is to demonstrate the general effect of a shell on phase separation. For a given material pair, faceting may or may not play a role in phase separation and this effect should be further investigated. In this study, we do not include the effect of plastic deformations since they do not happen at the length scale of interest based on both theoretical predictions [13] and experimental observations [14-16].

The model assumes a rotational symmetry along the axis of the nanowire. The circumferential component ($u_\theta$) of the displacement field is zero while the radial ($u_r$) and the axial ($u_z$) components of the displacement field are treated as variables. Displacement and traction vector continuity is assumed at all internal boundaries. The substrate is fixed (no displacements along $r$ and $z$ directions) at the bottom while all the external boundaries are traction-free. Semiconductor nanowires have been grown experimentally with diameters as small as 3 nm [17]. However, the range of diameters of interest is usually between 15-100 nm [15,18,19] due to difficulties associated with the growth of very small nanowires (<10 nm in diameter) and no practical advantage of larger nanowires (>100 nm in diameter) [17]. Throughout this study, the height of the nanowire and the diameter of the core are kept constant and equal to 200 nm and 20 nm, respectively, while the shell material and thickness ($t_s$) vary. GaAsSb is chosen as an example of immiscible alloy for the core nanowire. GaAsSb is thermodynamically unstable at 650 °C and under equilibrium condition, it phase separates to GaAs-rich and GaSb-rich phases. Given that this phase separation process is effectively determined by interdiffusion of As and Sb on the same sublattice, diffusion of Ga does not play an important role and hence we have not included the diffusivity of Ga in this model. Hence we use a single concentration variable in our model. We use the effective diffusion coefficient with the value of $1\times10^{-18}$ cm$^2$ s$^{-1}$ [20], which was determined in experimental studies of Sb diffusion in GaAs. We define free energy density functional as

$$F = \int_\Omega \left( f(c) + W(c) + \frac{\varepsilon^2}{2} |\nabla c|^2 \right) d\Omega \tag{1}$$

where $\Omega$ represents the system volume, $W$ is the strain energy density, $\varepsilon^2$ is the gradient energy coefficient, $f$ is the free energy density, and $c$ is the concentration of Sb. For GaAsSb at 650 °C, the excess Gibbs free energy is taken from CALPHAD calculations [21] and it is given by

$$f_{core} = \frac{(G_{ideal} + L^0_{GaAsSb} c(1-c) + L^1_{GaAsSb} c(1-c)(1-2c))}{V_m^{GaAs}} \tag{2}$$

where $V_m^{GaAs}$ is the molar volume of GaAs and $G_{ideal}$ the ideal Gibbs free energy. The interaction parameters $L^0_{GaAsSb}$ and $L^1_{GaAsSb}$ are defined as follows

$$G_{ideal} = \text{RT}(c \ln(c) + (1-c) \ln(1-c)) \tag{3}$$
$$L^0_{GaAsSb} = 24824 - 7.74301 \times \text{T} \tag{4}$$
$$L^1_{GaAsSb} = 4774 \tag{5}$$

where $R$ is the gas constant, and $T$ is the temperature in Kelvin. For numerical reasons, we fitted the Gibbs free energy with a ninth order polynomial (shown in Fig. 1) and we used this polynomial function in our model. The polynomial function is given by

$$G_{ideal} = -3.14808 \times 10^6 - 4.81905 \times 10^8 c + 5.02091 \times 10^9 c^2 - 2.64836 \times 10^{10} c^3 + 8.04754 \times 10^{10} c^4 - 1.4885 \times 10^{11} c^5 + 1.63584 \times 10^{11} c^6 - 9.76861 \times 10^{10} c^7 + 2.44215 \times 10^{10} c^8 + 2.01987 c^9 \tag{6}$$

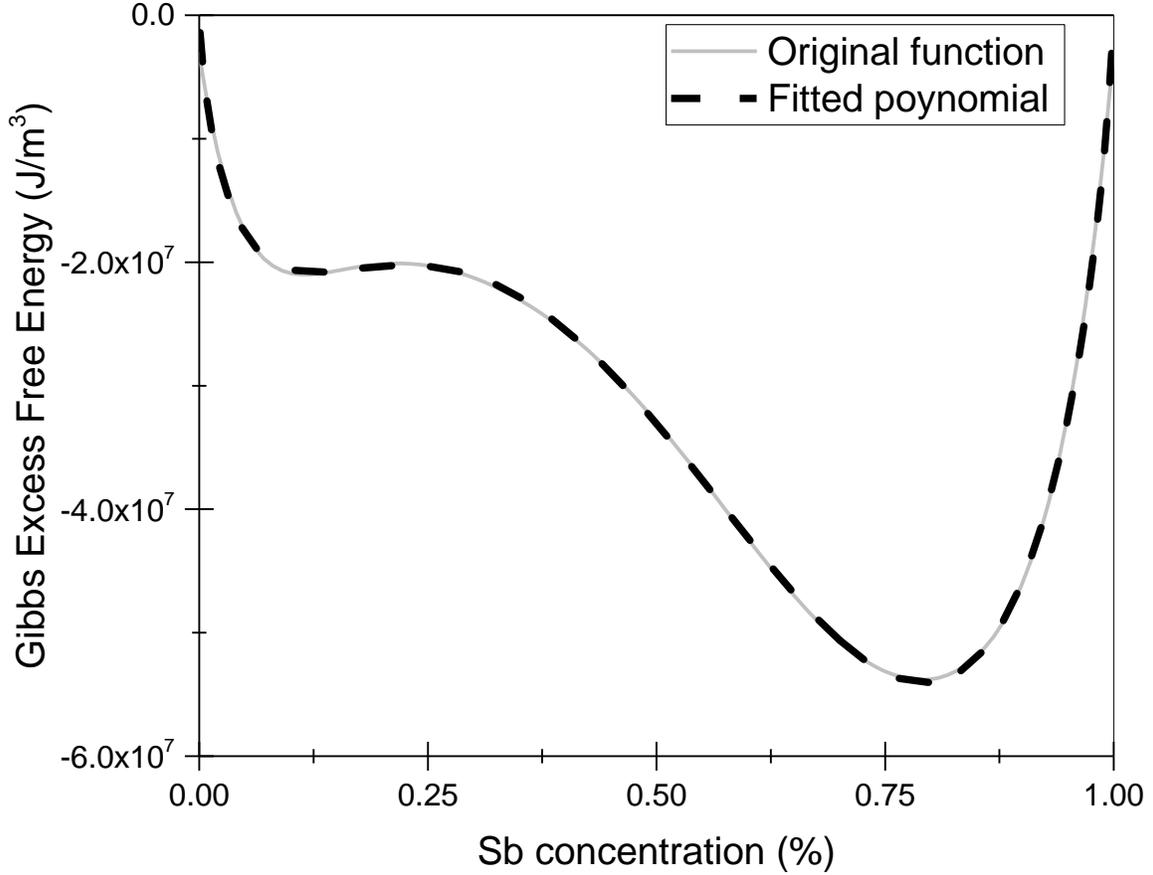

Fig. 1 Excess Gibbs free energy of GaAsSb at 650 °C (solid line). Ninth order polynomial has been used to fit the function (dashed line).

For the shell, we have chosen a material (GaAs) that is thermodynamically stable material at the temperature of interest (650 °C) and hence in our formulation it is sufficient to use a single-welled free energy density [22] as follows

$$f_{shell/substrate} = \alpha_0 c^2 \tag{7}$$

where $\alpha_0$ is a positive coefficient representing the sharpness of the single-welled function.

The second term in free energy functional shown in Eq. (1) is strain energy density $W$ that is defined as

$$W(c) = \frac{1}{2} \sigma_{ij} e_{ij}^{el} \tag{8}$$

$\sigma_{ij}$ is the Cauchy stress tensor, and $e_{ij}^{el}$ is the elastic strain tensor. Elastic strain tensor satisfies the following relationship

$$e_{ij}^{el} = e_{ij}^{tot} - e_{ij}^{*} \tag{9}$$

where $e_{ij}^{tot}$ is the total strain and $e_{ij}^{*}$ is the eigenstrain, which arises due to the lattice mismatch between core/shell and the substrate. $e_{ij}^{*}$ is given by

$$e_{ij}^{*} = \frac{a_{core/shell} - a_{substrate}}{a_{substrate}} \delta_{ij} \tag{10}$$

where $a$ is the lattice parameter of different parts of the nanowire specified in the subscript and $\delta_{ij}$ is the Kronecker delta. We use a linear interpolation to calculate the lattice mismatch between the core/the shell and the substrate as a function of alloy concentration. Assuming a linear strain-displacement relationship (which is valid for an infinitesimal strain), we solve equilibrium equations to find the total strain. Both the strain-displacement and the equilibrium equations can be found in Ref. [11].

The third term in free energy functional shown in Eq. (1) is interfacial energy that is a function of $\varepsilon^2$ and $c$. In our model, $c$ is a conserved field variable that evolves according to the mass conservation equation

$$\frac{\partial c}{\partial t} = -\nabla \cdot J \qquad (11)$$

where $J$ is the density flux and it is related to the gradient of variational derivative of the free energy density functional as

$$J = -M\nabla \frac{\delta F}{\delta c} = -M\nabla \left( \frac{\partial f}{\partial c} + \frac{\partial W^{el}}{\partial c} - \nabla \cdot (\varepsilon^2 \nabla c) \right) \qquad (12)$$

Here $M$ is the effective mobility of alloy. Mobility and diffusivity are related to each other through $MRT=DV_m$ [23], where $D$ is diffusivity and $V_m$ is the molar volume. There are two separate mass transport mechanisms included in our model, which are the surface and the bulk diffusion. In this model, the diffusion coefficient changes from $1\times10^{-8}$ cm$^2$ s$^{-1}$ at the surface (in the region corresponding to the thickness of 3 monolayers) to $1\times10^{-18}$ cm$^2$ s$^{-1}$ in the bulk. The thickness of the surface region is in agreement with the thickness of 2-4 monolayers reported in the literature based on experimental observations [24]. Combining Eqs. (11) and (12), the evolution equation, known as the Cahn-Hilliard equation, can be found

$$\frac{\partial c}{\partial t} = \nabla \cdot \left[ M\nabla \left( \frac{\partial f}{\partial c} + \frac{\partial W^{el}}{\partial c} - \nabla \cdot (\varepsilon^2 \nabla c) \right) \right] \qquad (13)$$

Eqs. (1) and (13) show that the strain energy is coupled to the phase filed model. This coupling means that in order to find the unknown variables in the model, we need to simultaneously solve for the Cauchy-Navier (equilibrium) equations in both the radial and the axial directions and the Cahn-Hilliard equation. Here the unknowns are the two displacement fields ($u_r$ and $u_z$) and the concentration field ($c$). All energy values in the model are defined per unit of length. The aforementioned system of partial differential equations is solved using the finite element method as implemented in the COMSOL software. In the finite element mesh, we used 10 grid points per unit of length (1 nm), which means that each grid has dimensions of 1 Å by 1 Å.

A schematic view of the nanowire structure is shown in Fig. 2a. The model consists of a substrate, a cap and a core. The shell will be shown in subsequent figures. The substrate is made of GaAs and the shell is made either of material GaAs or a composition of alloy GaAsSb to produce compressive or tensile strain in the core, respectively. The cap is made of a single element material and is chosen to be lattice matched with the core-nanowire. The results presented in this paper are not unique to a single semiconductor alloy and can be applied to alloys that have a miscibility gap at high temperatures (<700 °C) and that have a relatively high bulk diffusivity in the temperature range of interest. There are multiple phenomena that affect the phase separation in the core-shell nanowire. We isolated these phenomena to study the effect of each on the phase separation. In order to study the effects of misfit strain between core and shell and also shell thickness on phase separation in the core, we first assumed that there is no mass transport (i.e., atomic mixing) between different sections of the nanowire (substrate, cap, core, and shell). We then include the possibility of intermixing to elucidate the effect of mass transport between core and shell on phase separation in the core.

The material in the core is GaAs$_{0.6}$Sb$_{0.4}$, which separates into two phases under equilibrium conditions at 650 °C. GaAs-rich alloy (GaAs$_{0.92}$GaSb$_{0.08}$) and GaSb-rich alloy (GaAs$_{0.27}$

GaSb$_{0.73}$). This composition of GaAsSb is chosen because it is in the middle of a tie line of the phase diagram at the temperature of interest. Material properties of GaAs$_{1-x}$Sb$_x$ are calculated using a linear interpolation between properties of GaAs (x=0) and GaSb (x=1). For GaAs we take the following values for material properties [11]: elastic constants $C_{11}$=118.8 GPa, $C_{12}$=53.4 GPa, and $C_{44}$=59.6 GPa, surface diffusion coefficient equal to 1×10$^{-8}$ cm$^2$ s$^{-1}$. GaSb is assumed to have the following properties [25]: elastic constants $C_{11}$=83.3 GPa, $C_{12}$=40.2 GPa, and $C_{44}$=43.2 GPa, misfit strain with respect to GaAs equal to 7.83%. Bulk diffusion coefficient for GaAsSb is taken to be 1×10$^{-18}$ cm$^2$ s$^{-1}$ [20]. Surface diffusivity of GaSb is assumed to be equal to the surface diffusivity of GaAs. Annealing temperature is 650 °C. The interfacial energy $\gamma$ between GaAs and GaSb is 0.08 J/m$^2$. This value is calculated from the Young's equation knowing the surface energies of GaAs to be equal to 0.71 J/m$^2$ [26] and GaSb which is 0.63 J/m$^2$ [27], knowing that GaAsSb starts to grow in a layer-by-layer mode on GaAs [28] and hence the contact angle is zero. Following Cahn and Hilliard [29], gradient energy coefficient $\varepsilon_0^2$ is calculated from $\gamma = \int_{0.077}^{0.73} \sqrt{2\varepsilon_0^2 f(c)}$ to be 6.52×10$^{-10}$ J/m. We normalize the simulation parameters by defining three characteristic units: $e^*$ is the characteristic energy chosen to be 10$^{10}$ J m$^{-3}$, $l^*$ is the characteristic length which is equal to 10$^{-9}$ m, and $t^*$ is the characteristic time taken to be 10$^{-8}$ s. Using these characteristic units, one can define dimensionless quantities for the free energy density ($f^* = \frac{f}{e^*}$), the elastic moduli ($C_{ij}^* = \frac{C_{ij}}{e^*}$), the strain energy density ($W^* = \frac{W}{e^*}$), the gradient energy coefficient ($\varepsilon^{*2} = \frac{\varepsilon^2}{l^{*2}.e^*}$) and the mobility ($M^* = \frac{M.t^*.e^*}{l^{*2}}$). Finally, the non-dimensional form of the Cahn-Hilliard equation is

$$\frac{\partial c}{\partial \hat{t}} = \nabla.\left[M^*\nabla\left(\frac{\partial f^*}{\partial c} + \frac{\partial W^{el*}}{\partial c} - \nabla.(\varepsilon^{*2}\nabla c)\right)\right] \quad (14)$$

where $\hat{t}$ is the computational time step and $\nabla$ represents gradient with respect to non-dimensional variables $\hat{x}$ and $\hat{y}$.

The results shown in this paper are generated using a 2D axisymmetric model of the nanowire because such model is computationally less expensive than a full 3D model. We have compared and validated the results of the 2D model against a full 3D model for selected cases and found the results to be similar. Specifically, the phase separation in both models starts at the free surface and then propagates inside the material. Also, in both models we observed that the phase separation is delayed when there is a thin shell around the core and by increasing the thickness of the shell the phase separation is suppressed. Comparison between 2D and 3D models is more qualitative since some of the parameters such as the interfacial energy coefficient have different values in 2D and 3D models and hence a quantitative comparison is difficult.

## 3. Results

We first use our model to understand the role of free surfaces in phase separation of semiconductor nanowires and therefore we simulate annealing of a core-only nanowire (i.e., shell thickness $t_s$=0). We begin the simulation with a homogenous GaAs$_{0.6}$Sb$_{0.4}$ nanowire on a GaAs substrate (Fig. 2a). After annealing for 7×10$^4$ s, compositional modulations develop at the free surface, as shown in Fig. 2b. At this point in time, only 60% of the alloy is phase separated (the cutoff for phase separation in this study is chosen to be ± 2.5% of the initial composition). There are two driving forces for phase separation. The first driving force is the relaxation of the stress (and the strain energy) near the substrate as the GaAs-rich phase newly formed at the bottom of the nanowire has a lower misfit strain with the substrate. The second driving force is the reduction in bulk free energy, since phase separation is thermodynamically favorable. Formation of LCM during annealing has been previously observed experimentally in III-V semiconductors by Hsieh *et al.* [9] and studied theoretically by Tang *et al.* [10]. After annealing for a longer time, our simulations reveal that phase separation propagates toward the center of the nanowire and, for instance, at $5\times10^5$ s about 93% of the original homogenous alloy is phase separated into GaAs-rich and GaSb-rich phases (Fig. 2c). Propagation of compositional modulation into the bulk is driven by relaxation of alloy's bulk free energy. Although, the strain energy cost for formation of axial heterostructures acts as a counterbalance force that slows down the propagation process, this energy cost is relatively low. Specifically, due to phase separation, strain energy increases by $0.106\times10^{-8}$ J while bulk free energy decreases by $1.495\times10^{-8}$ J (Fig. 2d). In addition, the difference between lattice parameters (and hence volumes) of GaAs-rich and GaSb-rich phases causes formation of undulations at the free surface, as shown in the magnified view in Fig. 2c. These undulations partially relax the epitaxial stress.

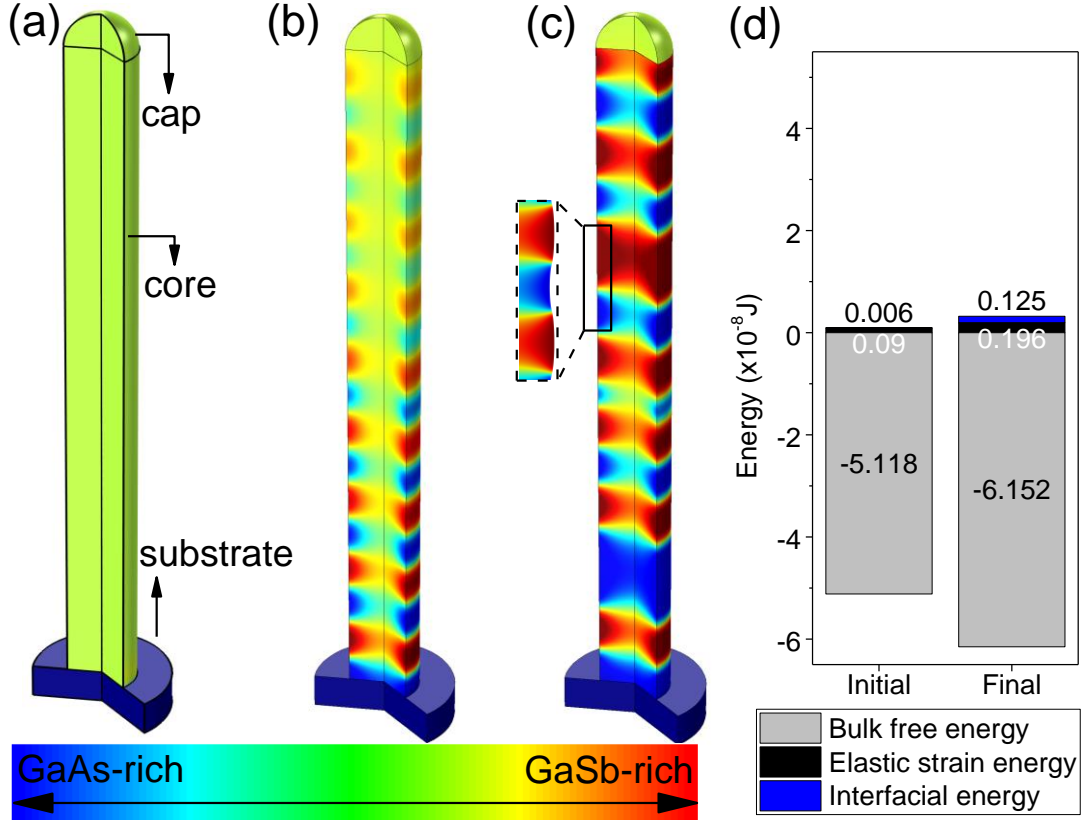

Fig. 2 GaAsSb core-only nanowire during annealing (a) before annealing, (b) after $7\times10^4$ s and (c) after $5\times10^5$ s of annealing at 650 ℃. Separated phases are GaSb-rich phase (red) and GaAs-rich phase (blue). (d) Contribution of different energy components before annealing and after annealing for $5\times10^5$ s.

Keeping in mind that phase separation is usually not desirable for opto-electronic applications of III-V semiconductors, we next focus on possible pathways for suppressing it. Specifically, we investigate whether including a shell around the nanowire's core can allow for control of phase separation in the regime where phase separation is thermodynamically favorable. To answer this question, we consider two specific cases. First, we include a thin shell ($t_s$=0.5 nm) made of GaAs around the core as shown in Fig. 3a. The 0.5 nm thickness corresponds to approximately 1 monolayer. Given that the lattice parameter of GaAs is smaller than GaAsSb, the shell imposes a compressive strain to the core. The purpose of having a thin shell around the core-nanowire is to enforce a kinetic constraint by substituting surface diffusivity with bulk diffusivity and therefore to determine the effect of surface diffusion on compositional modulation. Our results (Fig. 3) show that a thin shell only delays phase separation, but it does not suppress it entirely. This delay is due to the fact that a shell (even if it is thin) eliminates the fast surface diffusivity and as a result the kinetics is controlled by slower bulk diffusivity. Phase separation is still initiated at the core surface (which has now become the core-shell interface) and the same driving force as discussed for the core-only nanowire is active here. However, in the case of the thin-shell nanowire there is an additional force that opposes phase separation. This opposing force is associated with the misfit strain between core and the shell, but it is not large enough to suppress the separation. After annealing the nanowire for $7\times10^4$ s, 13% of the alloy is phase separated, which is significantly lower than 60% of phase separation observed in the nanowire that does not have a shell around it as shown in Fig. 2. After longer annealing time ($5\times10^5$ s), the phase separated region extends toward the center of the nanowire (Fig. 3a). At this point, 90% of phase separation is observed in nanowire that is comparable to 93% that was observed for only core nanowire. Contributions of different energy components to the total energy at the beginning of the annealing process and after $5\times10^5$ s are shown in Fig. 3c. Bulk free energy (grey) is the dominant term, but it decreases slightly over time.

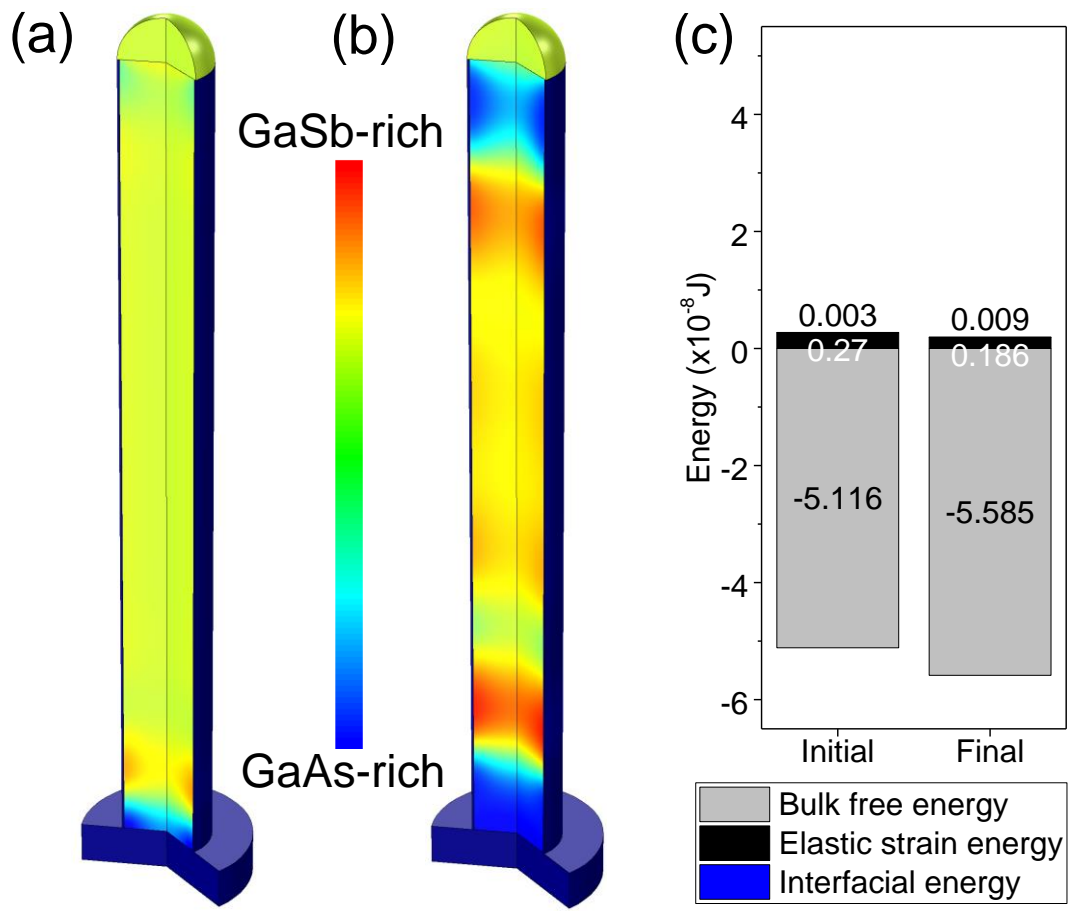

Fig. 3 (a) Composition of GaAsSb alloy core-shell nanowire with a GaAs thin shell after annealing for (a) $7\times10^4$ s and (b) $5\times10^5$ s. Separated phases are GaSb-rich phase (red) and GaAs-rich phase (blue). (c) Contributions of different energy components to the total energy before annealing and after annealing for $5\times10^5$ s.

We have shown that a thin shell around the core of a nanowire delays the phase separation. To explore whether such phase separation can be entirely suppressed we consider other geometries of the core-shell nanowire. In particular, it has been shown theoretically that stress and strain in the core increase when the shell thickness increases [12]. Here, we choose the shell thickness to be 5 nm as an example that illustrates our point. Similarly as in the thin shell model, GaAs is chosen as the shell and therefore the core is under the state of a compressive strain. After annealing at the same temperature of 650 °C as in the case of no-shell and thin-shell nanowires and for the same total amount of time (t = $5\times10^5$ s), we observe that LCM no longer develops ( Fig. 4a). Although some phase separation is observed near the core-substrate and core-cap interface, the phase separation after $5\times10^5$ s has been reduced to about 16% which is substantially smaller than phase separation in the core-only nanowire (93%) and the thin-shell nanowire (90%) after the same annealing time. We observe a very small change in the ratio of the elastic strain energy to the bulk free energy during annealing ( Fig. 4c). During annealing time, the total energy changes by less than 0.5%.

So far we discussed the effects of thin (0.5 nm) and thick (5 nm) shells on delay/suppression of the phase separation. In order to have a better understanding of the role of the shell thickness in phase separation, in Fig. 4d we report the results of annealing simulations for nanowires with a range of shell thicknesses. We found that in general when the shell thickness increases, the phase separation is suppressed but it does not decay all the way to zero. The reason for this behavior is that while the presence of the shell suppresses phase separation that starts from the core/shell interface, the phase separation can still be initiated at both ends of the nanowire. To demonstrate the effect of the nanowire ends on phase separation, in Fig. 4d we show the amount of phase separation in different sections of the nanowire. For instance, by removing 30% of the nanowire length, phase separation in the remaining 70% of the nanowire approaches 6% for a 10 nm shell. This suppression of phase separation in the mid-section of the nanowire can be attributed to the increase of the strain energy cost associated with lattice mismatch between the phase-separated core and the shell. This energy term increases as the shell thickness increases. The only other contribution to the cost in energy is the strain energy arising from the lattice mismatch between the layers in the phase-separated core, but this energy contribution is not expected to depend on the thickness of the shell and, as shown in Fig. 2, it is quite small. These results demonstrate that geometry and strain can be exploited in nanowire heterostructures as means for suppressing phase separation in an alloy with a thermodynamic driving force to separate at relatively high annealing temperatures. All the results in this paper have been generated for nanowires with diameters equal to 20 nm because of the computational efficiency. However, for selected cases we have also performed simulation for thinner (10 nm in diameter) and thicker (40 nm in diameter) nanowires and we have found the results to be qualitatively the same as in the case of the 20 nm nanowire. For instance, the effect of shell thickness on suppression of phase separation (similar to that shown in Fig. 4d) has been observed both in thinner and in thicker nanowires.

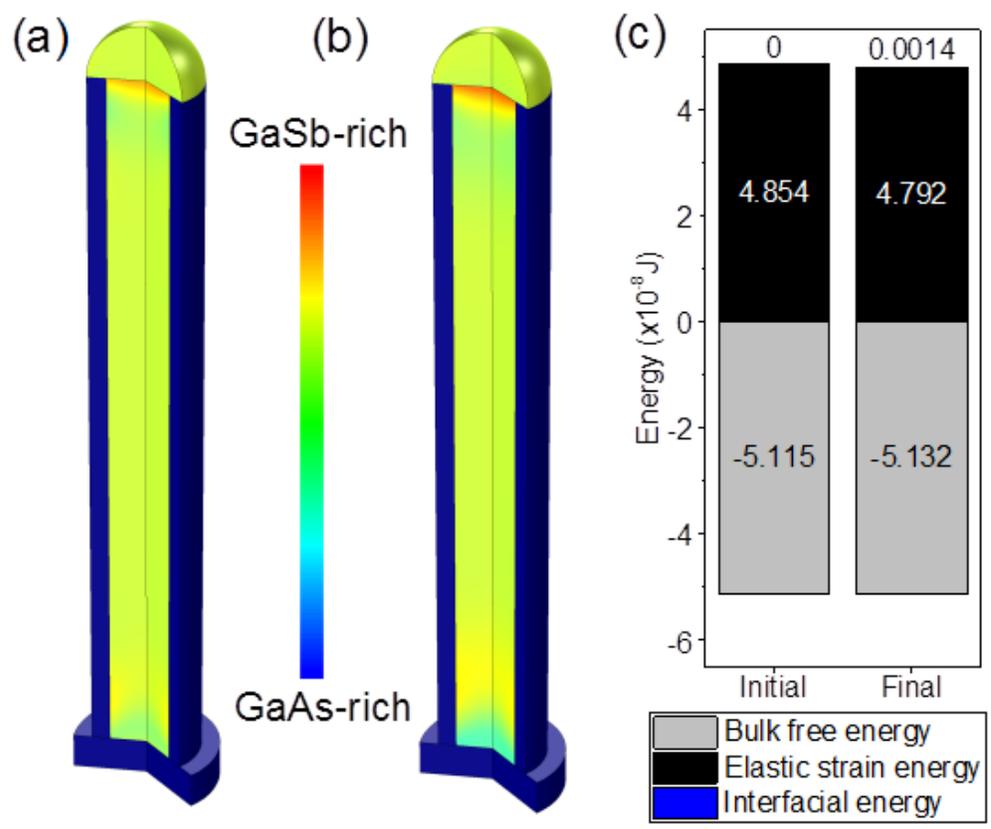
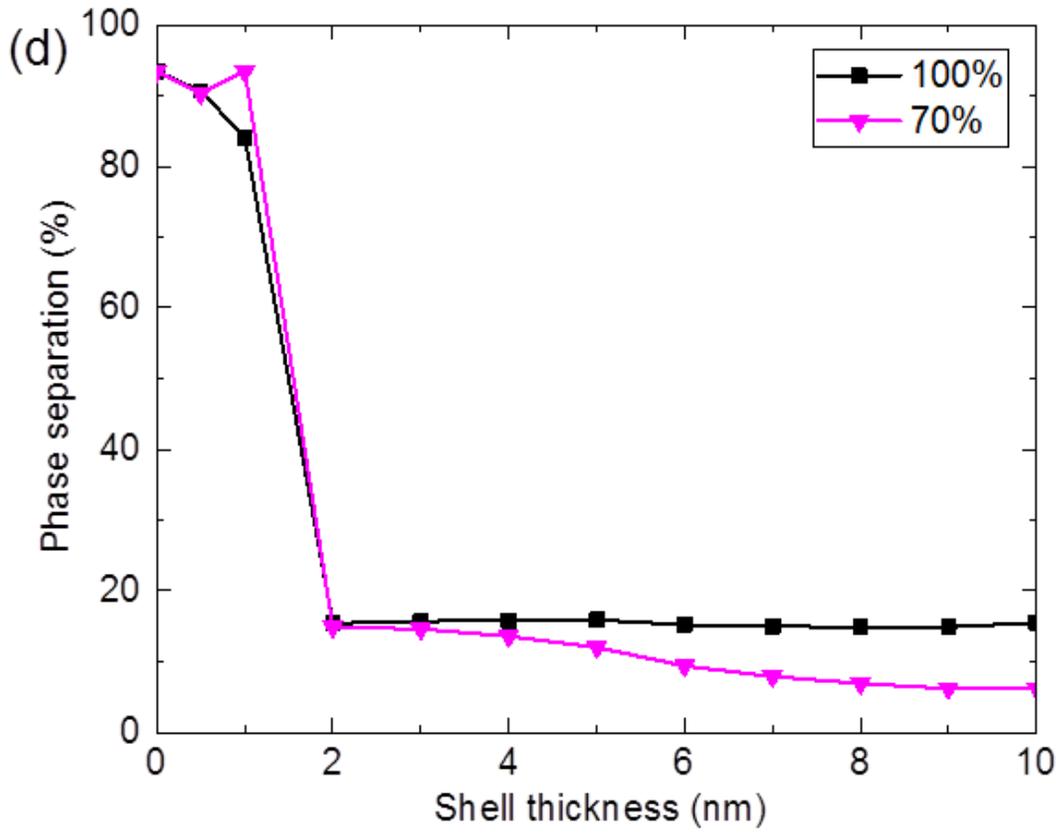

Fig. 4 (a) Composition of GaAsSb core-shell nanowire with GaAs thick shell after annealing for (a) $7\times10^4$ s and (b) $5\times10^5$ s. Separated phases are GaSb-rich phase (red) and GaAs-rich phase (blue). (c) Contribution of different energy components before annealing and after annealing for $5\times10^5$ s. (d) Effect of shell thickness on suppression of phase separation. The thicker the shell, the lower the phase separation.

Up until now we have considered the effects of a compressive strain imposed by the GaAs shell (GaAs has -3.13% misfit strain with the $GaAs_{0.6}Sb_{0.4}$ core before annealing). It is interesting to investigate the effect of misfit strain between core and shell on phase separation in the core. For this reason we model high temperature annealing in a nanowire with shells made of GaSb (+4.69% misfit strain with $GaAs_{0.6}Sb_{0.4}$ core before annealing), $GaAs_{0.28}Sb_{0.73}$ (+2.58% misfit strain with the $GaAs_{0.6}Sb_{0.4}$ core before annealing), $GaAs_{0.6}Sb_{0.4}$ (lattice matched with the GaAsSb core before annealing) and $GaAs_{0.922}Sb_{0.078}$ (-2.53% misfit strain with the $GaAs_{0.6}Sb_{0.4}$ core before annealing). Comparison of the amount of phase separation under compressive, no misfit strain, and tensile strain during the same period of annealing ($5\times10^5$ s) is shown in Fig. 5a. Our results show that compressive strain suppresses the phase separation more than tensile strain. However for a case where there is a compressive misfit strain between core and shell (-2.53%), the model predicts that the phase separation is minimized. The high cost of strain energy between separated phases and the compressive $GaAs_{0.922}Sb_{0.078}$ shell compared to low gain in bulk free energy during phase separation is the reason for the suppression of such separation. Although the strain energy cost due to the lattice mismatch between the core and the shell cannot be easily isolated in our simulations, it must be lower when the core is under tensile stress (GaSb and $GaAs_{0.28}Sb_{0.73}$ shells) than when it is under compression (GaAs and $GaAs_{0.922}Sb_{0.078}$ shell) or no strain ($GaAs_{0.6}Sb_{0.4}$) since more phase separation has been observed in these cases. In general, the exact amount of elastic strain energy arising from lattice mismatch between the core and the shell will depend on the specific thermodynamic phases to which the multicomponent alloy separates, on their lattice parameters and elastic constants, as well as on the properties of the shell.

Finally, we have investigated the effect of intermixing between the core and the shell on suppression of phase separation in the core. In Fig. 5b we compare the effect of shell thickness on phase separation of 5 nm GaAs shells with and without intermixing between the core and the shell after annealing for $5\times10^5$ s (material properties such as elastic constants, free energies, and diffusivities are kept constant). Similarly as in the case of when no intermixing is allowed, phase separation is significantly suppressed. However, the suppression of phase separation is less pronounced when intermixing is allowed. The effect of shell thickness on phase separation in the two cases is similar when the shell thickness is small; in this regime phase separation decreases with an increasing shell thickness. When the shell thickness is sufficiently large (here larger than 3 nm), phase separation saturates for the no-intermixing case (due to the end effects of a nanowire as discussed earlier) and it increases slightly with shell thickness for the case when intermixing is allowed. The reason for the latter trend is that strain energy in the core-shell structure increases as a function of the shell thickness and intermixing can relax this strain energy at the cost of increasing the bulk free energy in both the core and the shell. One should note that the intermixing occurs only in a relatively small fraction of the nanowire at the interface between the core and the shell (this intermixed region has the thickness of 2.0-2.5 nm).

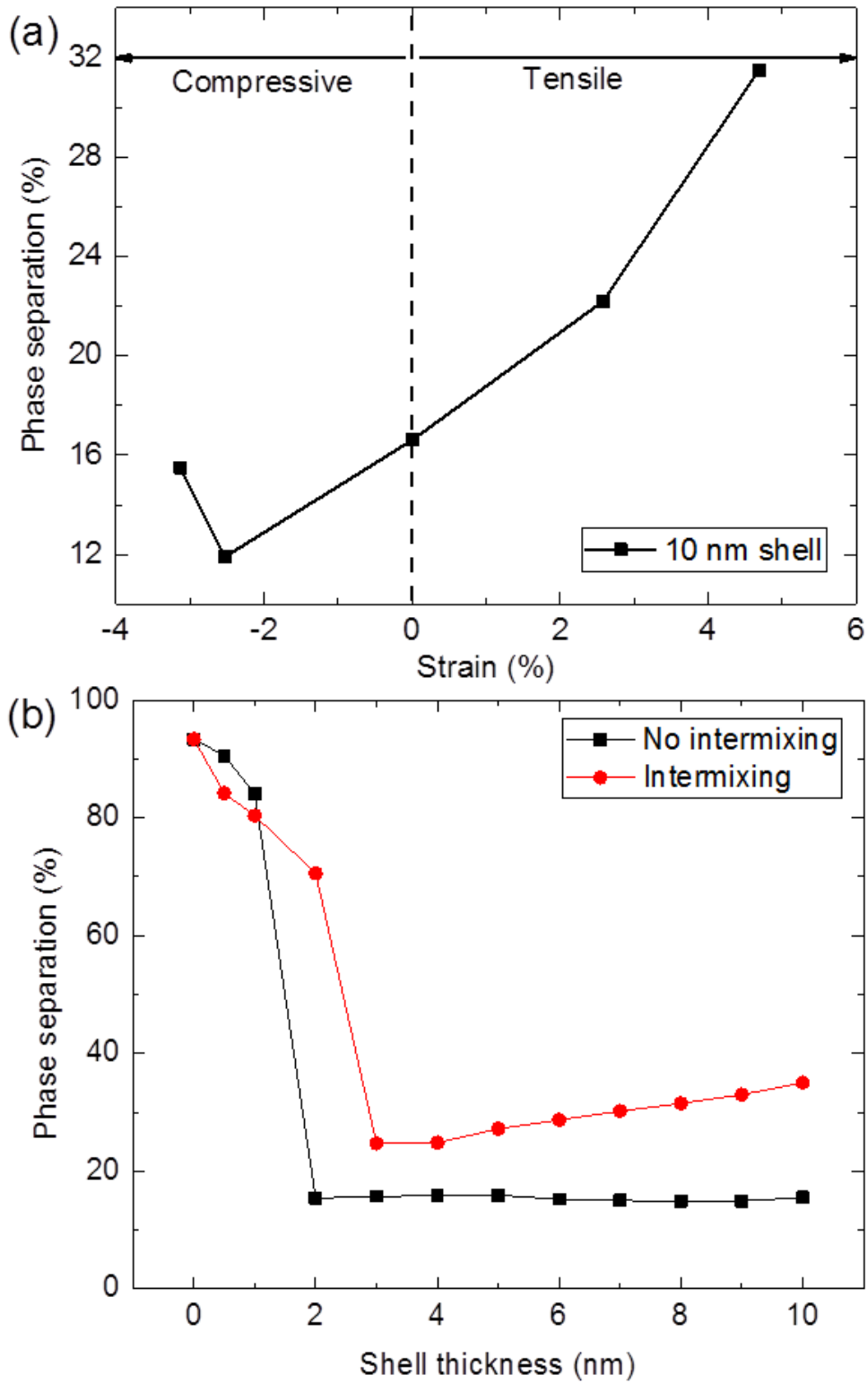

Fig. 5 (a) Effect of misfit strain between the core and the shell on phase separation in the core. Compressive strain suppreses phase separation more than tensile strain. Maximum suppression of phase separation happens for -2.53% misfit strain. (b) Comparision between the effect of shell thickness on phase separation with and without intermixing allowed between the core and the shell. Intermixing increases the amount of phase separation taking place in the core when shells are sufficiently thick, but even in that case the shell is still able to suppress a significant amount of phase separation.

## 4. Summary and conclusions

We demonstrated that during annealing, a core-only nanowire develops surface-induced compositional modulations that take the form of axial heterostructures. Including a shell around the core-nanowire controls phase separation by two mechanisms. The first mechanism involves kinetics where phase separation is suppressed by removing surface diffusion as a pathway for mass transport. The second mechanism involves thermodynamics where the thicker the shell, the higher the elastic energy cost associated with lattice mismatch between the separated phases of the core and the shell. Both lattice matched and lattice mismatched shells suppress phase separation in the core. However, the effect is not necessarily symmetric and depends on the difference between lattice constants on the phases formed in the core during phase separation and the material of the shell.

## Acknowledgements

This research was primarily supported by University of Wisconsin Materials Research Science and Engineering Center (DMR-1121288).

## References

[1]     R. X. Yan, D. Gargas, and P. D. Yang, Nat Photonics **3**, 569 (2009).
[2]     X. F. Duan, Y. Huang, Y. Cui, J. F. Wang, and C. M. Lieber, Nature **409**, 66 (2001).
[3]     X. F. Duan, C. M. Niu, V. Sahi, J. Chen, J. W. Parce, S. Empedocles, and J. L. Goldman, Nature **425**, 274 (2003).
[4]     M. C. McAlpine, H. Ahmad, D. W. Wang, and J. R. Heath, Nat Mater **6**, 379 (2007).
[5]     Q. Gao, H. H. Tan, H. E. Jackson, L. M. Smith, J. M. Yarrison-Rice, J. Zou, and C. Jagadish, Semicond. Sci. Technol. **26**, 014035 (2011).
[6]     Y. N. Guo *et al.*, Nano Lett. **13**, 643 (2013).
[7]     C. A. Wang, J. Electron. Mater. **29**, 112 (2000).
[8]     E. Luna, M. Wu, J. Puustinen, M. Guina, and A. Trampert, J. Appl. Phys. **117**, 185302 (2015).
[9]     K. C. Hsieh, K. Y. Hsieh, Y. L. Hwang, T. Zhang, and R. M. Kolbas, Appl. Phys. Lett. **68**, 1790 (1996).
[10]    M. Tang and A. Karma, Phys. Rev. Lett. **108**, 265701 (2012).
[11]    M. Arjmand, J. Deng, N. Swaminathan, D. Morgan, and I. Szlufarska, J. Appl. Phys. **116**, 114313 (2014).
[12]    Y. Liang, W. D. Nix, P. B. Griffin, and J. D. Plummer, J. Appl. Phys. **97**, 043519 (2005).
[13]    O. Salehzadeh, K. L. Kavanagh, and S. P. Watkins, J. Appl. Phys. **113** (2013).
[14]    K. L. Kavanagh, Semicond. Sci. Technol. **25**, 024006 (2010).
[15]    L. J. Lauhon, M. S. Gudiksen, C. L. Wang, and C. M. Lieber, Nature **420**, 57 (2002).
[16]    K. M. Varahramyan, D. Ferrer, E. Tutuc, and S. K. Banerjee, Appl. Phys. Lett. **95**, 033101 (2009).
[17]    L. Chen, W. Lu, and C. M. Lieber, in *Semiconductor Nanowires: From Next-Generation Electronics to Sustainable Energy* (The Royal Society of Chemistry, 2015), pp. 1.
[18]    H. L. Zhou, T. B. Hoang, D. L. Dheeraj, A. T. J. van Helvoort, L. Liu, J. C. Harmand, B. O. Fimland, and H. Weman, Nanotechnology **20**, 415701 (2009).
[19]    D. W. Wang, R. Tu, L. Zhang, and H. J. Dai, Angew Chem Int Edit **44**, 2925 (2005).
[20]    O. M. Khreis, K. P. Homewood, W. P. Gillin, and K. E. Singer, J. Appl. Phys. **84**, 4017 (1998).
[21]    K. Ishida, T. Shumiya, T. Nomura, H. Ohtani, and T. Nishizawa, J Less-Common Met **142**, 135 (1988).

[22] A. Boyne, M. D. Rauscher, S. A. Dregia, and Y. Wang, Scripta Mater. **64**, 705 (2011).
[23] N. Moelans, B. Blanpain, and P. Wollants, Computer Coupling of Phase Diagrams and Thermochemistry **32**, 268 (2008).
[24] Y. Tu and J. Tersoff, Phys. Rev. Lett. **98**, 096103 (2007).
[25] W. F. Boyle and R. J. Sladek, Phys Rev B **11**, 2933 (1975).
[26] E. Placidi *et al.*, in *Self-Assembly of Nanostructures: The INFN Lectures*, edited by Springer2012).
[27] G. Guisbiers, M. Wautelet, and L. Buchaillot, Phys Rev B **79** (2009).
[28] R. T. Hao, Y. Q. Xu, Z. Q. Zhou, Z. W. Ren, H. Q. Ni, Z. H. He, and Z. C. Niu, J Phys D Appl Phys **40**, 1080 (2007).
[29] J. W. Cahn and J. E. Hilliard, J. Chem. Phys. **28**, 258 (1958).